\documentclass[a4paper, conference]{IEEEtran}
\ifCLASSINFOpdf
  \usepackage[pdftex]{graphicx}
  \graphicspath{Graphics}
  \DeclareGraphicsExtensions{.pdf}
\else
\fi
\usepackage[font=footnotesize]{subfig}
\usepackage{url}

\usepackage{color}

\usepackage{listings}
\lstdefinestyle{tech}{
  basicstyle=\footnotesize\ttfamily,
  breaklines=true,
  breakatwhitespace=true
}

\newcommand{\eg}{\textit{e.g.,\ }}
\newcommand{\etc}{\textit{etc.\ }}
\newcommand{\ie}{\textit{i.e.,\ }}
\newcommand{\etal}{\textit{et al.\ }}

\begin{document}
%
\title{Architectural Analysis of FPGA Technology Impact}

\author{\IEEEauthorblockN{Oriol Arcas-Abella, PhD}
\and
\IEEEauthorblockN{Abhinav Agarwal, PhD}
}



\maketitle

\begin{abstract}
The use of high-level languages for designing hardware is gaining popularity since they increase design productivity by providing higher abstractions.
However, one drawback of such abstraction level has been the difficulty of relating the low-level implementation problems back to the original high-level design, which is paramount for architectural optimization.
In this work\footnote{This project was started in April 2013.}, we propose a methodology to analyze the effects of technology over the architecture, and to generate architectural-level area, delay and power metrics. Such feedback allows the designer to quickly gauge the impact of architectural decisions on the quality of generated hardware and opens the door to automatic architectural analysis. We demonstrate the use of our technique on three FPGA platforms using two designs: a Reed-Solomon error correction decoder and a 32-bit pipelined processor implementation.

\end{abstract}


%
\IEEEpeerreviewmaketitle

\section{Introduction}

In recent years, the automation of digital hardware design has advanced greatly. The industry has concurrent goals of increased designer productivity, shorter design cycles and ambitious area-performance-power constraints. 
As a consequence designers are increasingly adopting high-level Hardware Description Languages (HDL)~\cite{Chisel12, Bluespec:www, canis2011legup, chen2005xpilot, villarreal2010designing}
to address the challenging design requirements of large systems without losing productivity. These languages provide a higher level of abstraction than traditional languages such as VHDL or Verilog, implicitly hiding some of the low-level details in favor of an explicit system-wide view.

Existing commercial and academic design tools are good at synthesis, placement and routing of hardware designs for different implementation platforms. Usually such tools provide very detailed reports, but only on gross and macro-level resource consumption and performance metrics. Even if the detailed reports are analyzed, the resources are closely tied to the implemented circuit. Therefore, it is difficult to obtain a breakdown of area, delay and power metrics for the modules and sub-modules in the high-level design source.

Architecture designers require tools to automatically generate such a breakdown, as typical design iterations in the design cycle are always limited to specific blocks and modules. FPGAs are an important part of the high-performance ecosystem, as accelerators in heterogeneous architectures or high-throughput custom machines. In both cases they require deep design-space exploration~\cite{Benson12} and architectural refinements~\cite{Moussalli13} to appropriately map onto the target technology with the required constraints. There is an urgent need to improve and enhance the feedback from downstream synthesis tools to inform high-level design decisions.

This work tries to reduce the gap between the target circuits and their high-level architectural descriptions. In addition, bridging these two levels of the design process is the first and necessary step towards automatic architectural optimization.

Our contributions are the following:
\begin{itemize}
\item We present a novel methodology to analyze high-level designs and their synthesis reports in order to generate module and sub-module level metrics for area, delay and power. This methodology can be used with existing languages and hardware synthesis tools, and some of its metrics cannot be obtained by the existing design tools. The novelty of this work lies in automatic generation of metrics corresponding to high-level architectural units.
\item To demonstrate the applicability of this methodology, we implement it on a rule-based HDL.
We use our prototype implementation to extract results for two micro-architectural test cases: a Reed-Solomon decoder and a RISC microprocessor. We demonstrate the use of our tool on three different FPGA platforms.
\item The results presented are enriched with architectural information. We propose how they could be used by the designer to take informed architectural decisions in order to optimize the design.
\item We also discuss the possibility of automatic architectural optimization using our framework.
\end{itemize}



\emph{Paper Organization:} Section \ref{sec:meth} introduces our methodology. In Section \ref{sec:results}, we apply the methodology to the chosen high-level hardware designs implemented on three FPGA platforms and show three different metrics for each implementation. We consider the possibility of automatic architectural optimization in Section~\ref{sec:archopt}. In Section \ref{sec:applicability}, we discuss the applicability of our technique to other HLS frameworks, and present recently developed techniques and tools related to this work. Finally, Section \ref{sec:conc} describes future research directions and concludes the paper.

\section{Methodology}
\label{sec:meth}

\newsavebox{\gcdlisting}
\begin{lrbox}{\gcdlisting}
\begin{minipage}[b]{0.33\textwidth}
\begin{lstlisting}[style=tech]
Reg#(32) x
Reg#(32) y

rule swap (x > y && y > 0)
  x <= y
  y <= x

rule subtract (x <= y && y > 0)
  y <= y - x
\end{lstlisting}
\end{minipage}
\end{lrbox}

\begin{figure*}[t!]
  \subfloat[Bluespec architectural model.\label{fg:gcd_dataflow}]{
    \usebox{\gcdlisting}
  }
  \subfloat[Annotated RTL model.\label{fg:gcd_circuit}]{
    \includegraphics[width=0.33\textwidth]{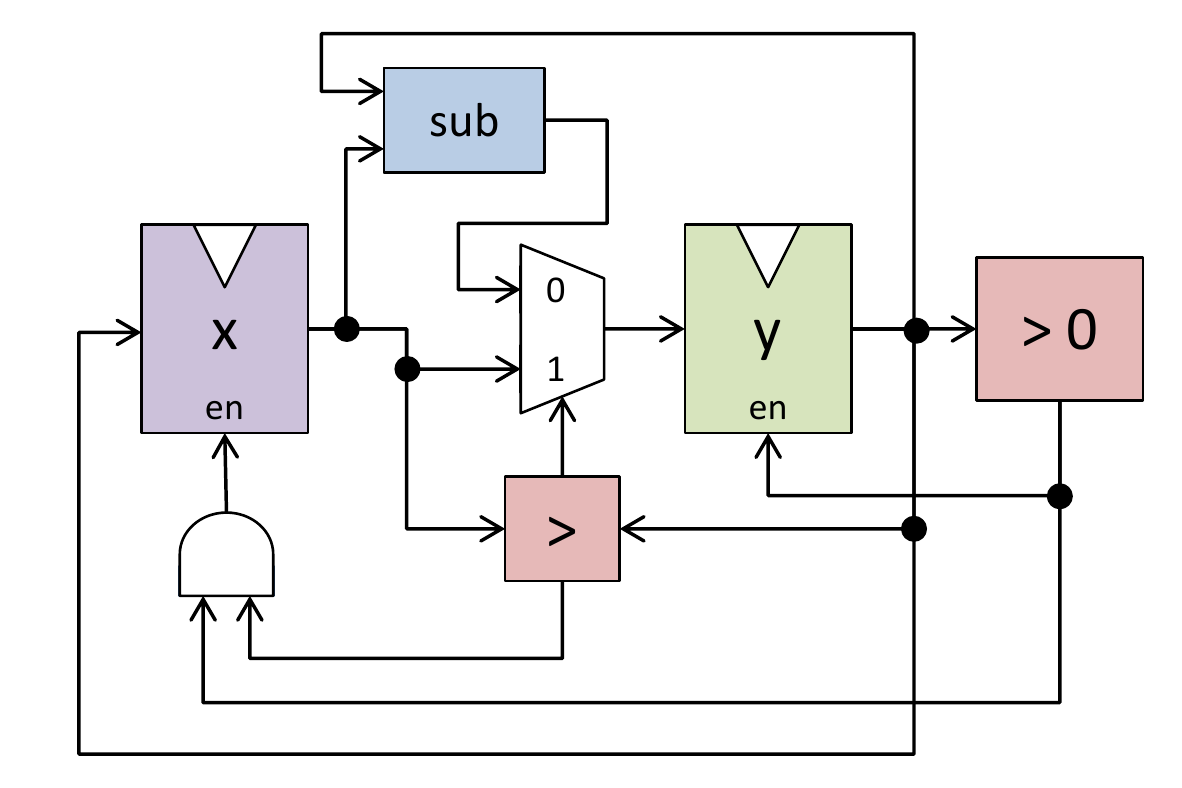}
  }
  \subfloat[Graph of the FPGA circuit.\label{fg:gcd_graph}]{
    \includegraphics[width=0.33\textwidth]{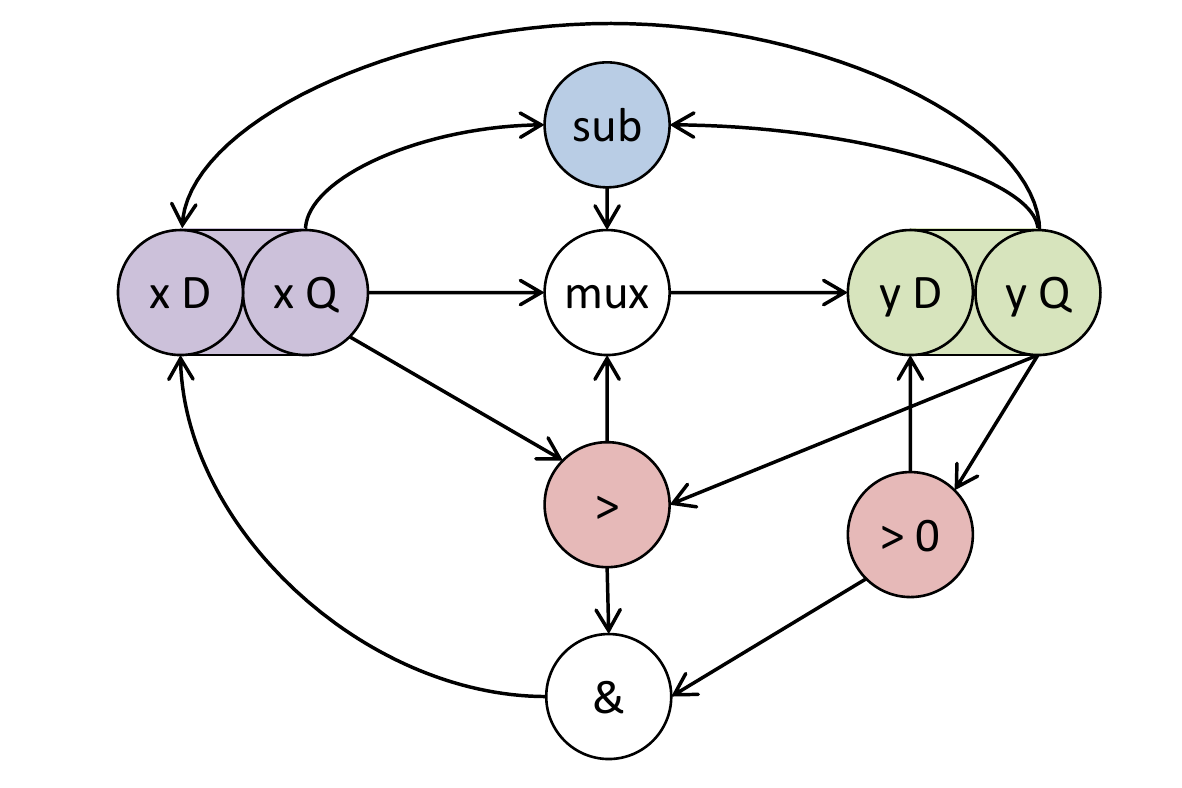}
  }
  \caption{Example of our methodology applied to a GCD module, from the architecture to annotated circuits.}\label{fg:gcd_example}
\end{figure*}

We divided our methodology in two steps. Firstly, the design is analyzed and annotated. Then, after passing through the synthesis flow, the resulting circuit is analyzed and compared to the original, abstract design. The user can apply architectural solutions to implementation problems, or even automatic architectural optimizations could be possible. Also, as the architectural and low-level information is known by the tool, automatic optimizations could be applied.

The analysis starts with the user's description of the architecture. As the needs for large-scale system design increase, several HDLs have been proposed.
To implement our methodology and demonstrate its applicability we have chosen Bluespec SystemVerilog~\cite{Bluespec:www}, a well-known rule-based HDL.
In section \ref{sec:applicability} we describe how our methodology can be applied to HLS tools as well.

As an example, consider the traditional swap/subtract greatest common divisor (GCD) Euclidean algorithm. In Figure \ref{fg:gcd_dataflow} we show a 32-bit GCD module implemented with two Bluespec rules.
This example contains two rules and two 32-bit registers. Both guards are mutually exclusive, so at most one rule can be executed every cycle.

\subsection{Analysis of the Architecture}
\label{sec:thesis}




The initial step of our methodology starts with a user description of a hardware architecture. We show the user description of the GCD design in Figure~\ref{fg:gcd_dataflow}.
Instead of directly analyzing the high-level source code, our tool analyzes the intermediate representation to simplify the problem. 



Our tool generates an abstract representation of the architecture from the Bluespec description.
During the rest of the paper, we term the high-level objects (submodules and rules) as ``blocks''. In Figure~\ref{fg:gcd_dataflow} there are four blocks: the rules \emph{swap} and \emph{subtract}, and the registers (submodules) \emph{x} and \emph{y}.
The intermediate representation contains a very simplified model of the operations of the rules. All the algorithmic descriptions are converted into a flat RTL representation.
The intermediate representation is converted into the definitive Verilog description of the hardware, which is functionally equivalent to the architectural description in Bluespec.
The main goal of our methodology is to establish relations between the final circuit and the original architecture. The core technique to accomplish this is the annotation method, where innocuous
annotations are added to the Verilog files generated by the Bluespec compiler. This annotation process is specially necessary for high-level synthesis, where the generated RTL code does not clearly reflect the original architectural units and the algorithmic semantics.

This extra information will remain during the FPGA circuit synthesis, and
consists of unique identifiers that are assigned to each Bluespec block. This prefix is added to the name of each Verilog element that is present in the intermediate model, identifying the block that owns it.
An RTL circuit of the GCD design is shown in Figure~\ref{fg:gcd_circuit}. The annotations are represented as colors. For example, the subtract arithmetic unit, in blue, corresponds to the \emph{subtract} rule. Similar units are merged by the Bluespec compiler and may not be identified properly (with a white background in the example). This represents less than 1\% of the hardware.


\subsection{Analysis of the Circuit}




\begin{figure*}[t]
 \centering
 \subfloat[FPGA circuit \label{fg:delay_alg_a}]{
  \includegraphics[width=0.15\textwidth]{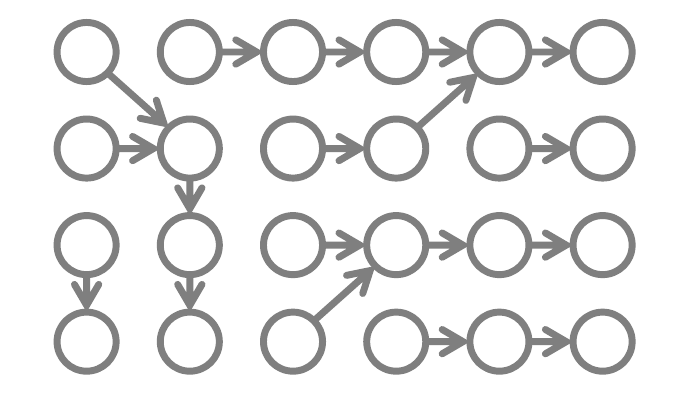}
 }
 \subfloat[Annotated nodes \label{fg:delay_alg_b}]{
  \includegraphics[width=0.15\textwidth]{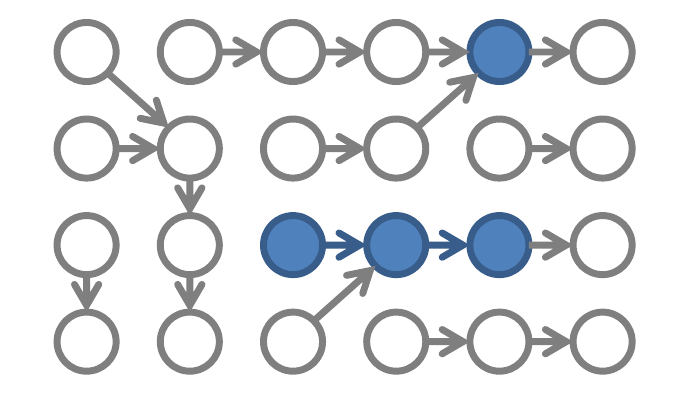}
 }
 \subfloat[Expanded paths \label{fg:delay_alg_c}]{
  \includegraphics[width=0.15\textwidth]{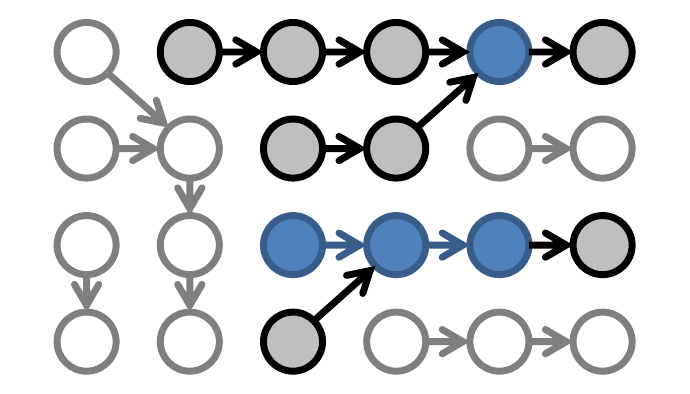}
 }
 \subfloat[Connected sets \label{fg:delay_alg_d}]{
  \includegraphics[width=0.15\textwidth]{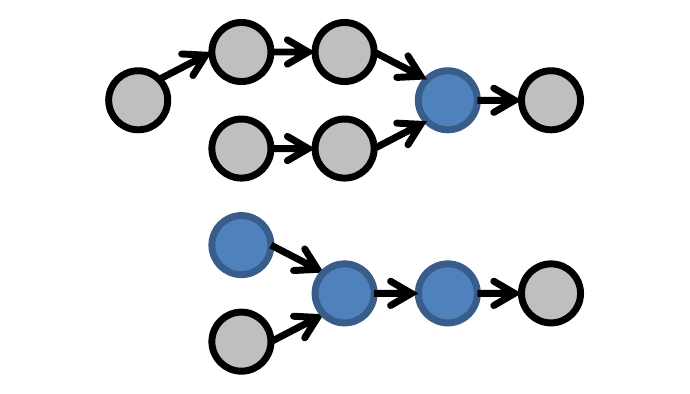}
 }
 \subfloat[System delays\label{fg:delay_alg_e}]{
  \includegraphics[width=0.15\textwidth]{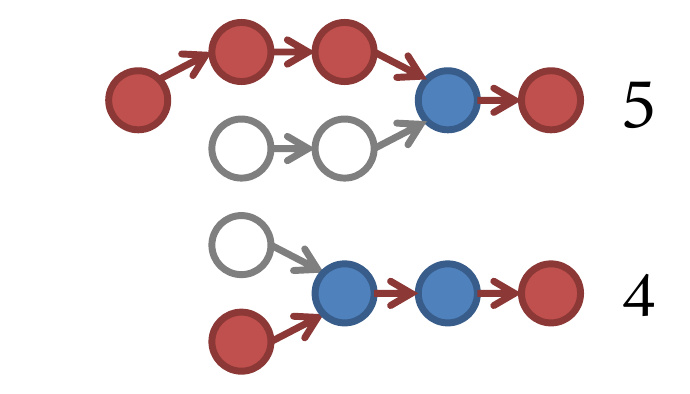}
 }
 \subfloat[Block delays \label{fg:delay_alg_f}]{
  \includegraphics[width=0.15\textwidth]{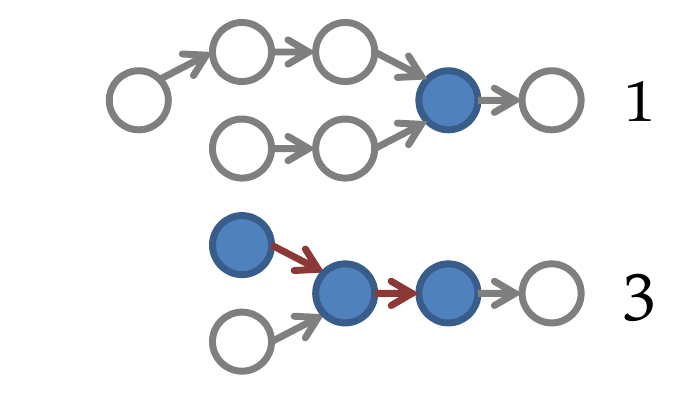}
 }
 \caption{Algorithm used to calculate the block delay. Nodes are FPGA cells, arrows are combinational paths.}
 \label{fg:delay_alg}
\end{figure*}

After the Verilog model is annotated, it gets synthesized into circuits (\ie mapped, placed and routed) using the FPGA vendor's tools. In this process the architecture is implemented in a particular technology, adapting to its characteristics.
The next step is analyzing this circuit and identifying the annotations that remain after the synthesis. This will enrich the available information on area, timing and power characteristics of the circuit, establishing a direct link with the original architecture.
During the following sections we will refer to the FPGA electronic elements,~\ie LUTs, registers,~\etc as cells or gates as well.

We show a hypothetical synthesis of our GCD model in Figure~\ref{fg:gcd_graph}. The two original registers have been synthesized into flip-flops. We show the input (D) and output (Q) ports of the flip-flops in separated nodes because there is no asynchronous connection between them. The other cells are LUTs with different configurations and number of inputs. Again, the annotations, shown in the example as colors, remain.

The resources and timing information extracted from the final circuit will be used to identify the annotated gates and map them to the original architectural element. In this work, some specialized elements like Block RAMs and DSP slices have been kept outside the scope of our analysis and will need further research.

\subsubsection{Area estimation}

The area analysis refers to the amount of FPGA resources used.
Our tool analyzes the final circuit and represents it as a Directed Acyclic Graph (DAG) for computation purposes.
The nodes of the graph are FPGA cells, and the edges are FPGA nets. As we explained previously, after synthesizing the circuit, the FPGA cells are named after the Verilog element that generated them. Our tool looks for prefixes in the names and extracts the unique label that identifies the original Bluespec block. 



\subsubsection{Delay estimation}
\label{sec:delay}
The delays of the circuit can be represented as weights in the DAG that the tool extracted. Figure~\ref{fg:gcd_graph} is an example of a DAG, where every node corresponds to a FPGA cell and every edge is a connection between cells.
Some of these graphs can be too big to analyze, even for a small design (\eg 10,000 cells). Our tool partitions the data and reduces the input set of the algorithms as much as possible to optimize the analysis time. 

The delay calculation proceeds as follows. Our tool obtains all the combinational paths that cross each block of the original architecture. 
The example DAG shown in Figure~\ref{fg:gcd_graph} has combinational paths from the output (Q) ports to the input (D) ports of the registers. 

In Figure~\ref{fg:delay_alg_a} we provide another DAG to describe the delay estimation algorithms more clearly. The tool performs a delay analysis at block level. Using the information collected during the area analysis, all the cells that belong to the current block are identified in the DAG. An initial subgraph is generated with all these nodes. Figure~\ref{fg:delay_alg_b} shows the annotated nodes in blue. This way,
all the possible combinational paths are identified.
In Figure~\ref{fg:delay_alg_c} we show how from the 4 original annotated nodes, two ``combinational trees'' (each with 2 branches) are identified with 7 and 5 nodes. The starting nodes of the paths are clock ports, and the end nodes are the inputs to registers or memories. The middle nodes are logic gates.

This expanded subgraph is then partitioned into connected sets. In Figure~\ref{fg:delay_alg_d} the two connected sets, with 7 and 5 nodes are represented. Every connected set is a directed acyclic subgraph without combinational connections to the other sets. Therefore it can be analyzed independently.
From all the connected sets, the longest delay is assigned to the block. In the delay report, for every block the critical path is displayed, effectively listing all the nodes.

We devised two weighting functions to obtain different delay metrics:
\begin{itemize}
\item \textbf{System delays:} all the edges are weighted with the corresponding delay. This weighting function makes the algorithm to choose the longest paths that contain one or more nodes of the current block.
\item \textbf{Block delays:} only the nodes that belong to the current block have a positive weight, while the rest have a null weight. With this weighting function, the algorithm will return the paths with the highest delay due to the current block.
\end{itemize}

The first weighting function allows us to find if a block contributes to the longest (critical) paths of the system, which are the performance bottlenecks of the whole design. This information can be returned by any typical timing analysis tool. The second weighting function allows us to find what we consider is the delay of a design block: among all the combinational paths that cross the block, the longest intra-block segment. This second metric helps the designer to understand the isolated delay of an architectural unit.

We want to note that the results of both metrics can be completely different, because a block's internal maximum path can be longer than the block's contribution to a system critical. \textbf{Therefore, non-critical optimization opportunities of architectural units (\ie block delays) are difficult to detect with typical timing analysis, and their effect can be higher than what would be expected~\cite{cornu2011hls}.}
For instance, in Figure~\ref{fg:delay_alg_e} the critical paths are shown in red. Using this weighting function, the designer can observe that the block in blue contributes to two system paths that have 5 and 4 nodes. Assuming that in this example all the delays are equal, the path with 5 nodes would be the longest path contributed by this block. Figure~\ref{fg:delay_alg_f} shows the result of applying the block contribution delay. Only the delays of the annotated nodes are considered. The paths are 1 and 3 nodes long. In this example the block (in blue) contributes to a critical path of the system with 5 nodes, but the critical delay of the block is 3 nodes long. Thus, the designer could choose a) to optimize the critical path of the system in order to reduce the maximum period and increase the frequency.
Another choice could be b) to optimize the delay of the block if it is a functionality used often, or even c) using a different, higher speed clock for it.

\subsubsection{Power score}
\label{sec:energy}


As part of the feedback generated by the tool, the component blocks of a design are ranked in the order of their average power consumption. The various parts of power score characterization are individually considered in the following aspects:

The static power consumption is directly proportional to the area of the component blocks, weighted by values for varying sizes of LUTs and flip-flops.
The dynamic power consumption of a block is directly proportional to the product of the capacitance of the switching elements and the frequency of transitions of block elements. Dynamic activity within the block can either occur when the rule corresponding to the block fires or when the input state for the rule changes as a result of another rule firing. We profile the design to obtain the number of event transitions as measured by the rule-based activity, as well as to obtain the relationship between output state affected by a rule and input state dependence for each rule.

State change in a given cycle leads to dynamic activity in the next cycle in each block, which depends on the changed state. In addition, rules that fire in a given cycle write their state changes in the same cycle, again consuming power. We assume that, whenever a state is updated, its value changes, leading to dynamic transitions in dependent blocks. The dynamic power computed for the design takes into account both of these described components.





The static and dynamic power for each block ($P_S$ and $P_D$ respectively) is determined using the cell characterization obtained in the area analysis and the FPGA cell power model described in ~\cite{Li05powermodeling}. We then use dependency analysis and profiled rule-firing statistics, to compute the average switching factor ($\alpha$) for each block of the design. This is then used to compute the average power consumption of each block given by equation~\ref{eq1}, where $f$ is the frequency of operation.

\begin{equation} P_{avg} = P_s + P_D \cdot \alpha \cdot f\label{eq1}\end{equation}

This metric is termed as ``power score''.
Though it is not a direct estimate of the exact power consumption due to assumption of upper-bound block-level activity, this metric can still help the designer to compare the relative amount of dynamic and static power of various blocks of the hardware. In addition, the probable hot spots of the architecture can be identified.
The value of the metric cannot be used to directly compare the static and dynamic power of a block implemented in the different FPGA technologies as we use the same power model in each case. The main value of the score is that the relative power distribution of the design in each FPGA platform will follow the same pattern as the power score for that platform, and this information can be used to analyze and reduce the power consumption in a customized manner.

\section{Results}
\label{sec:results}



\begin{figure}[t]
  \centering
  \subfloat[\label{fg:rs_arch} Reed-Solomon decoder]{
    \includegraphics[width=0.35\columnwidth]{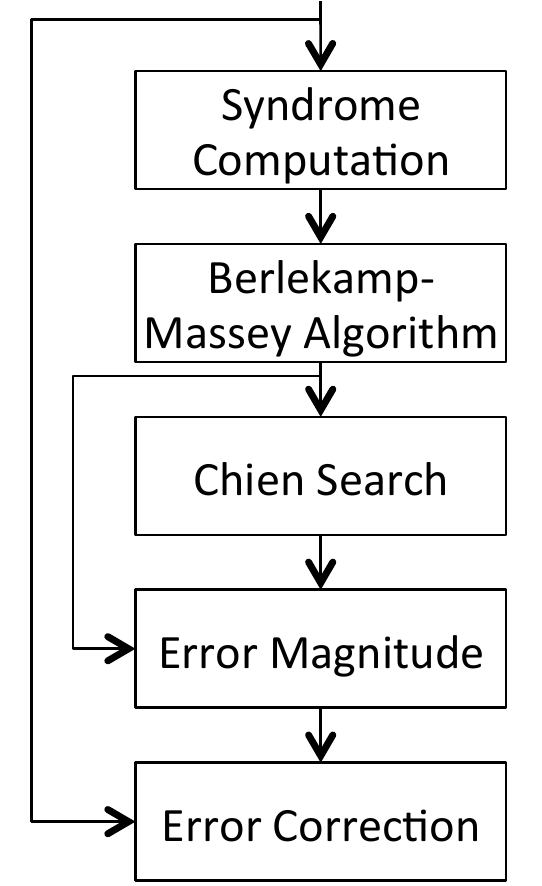}
  }
  \subfloat[\label{fg:smips_arch} SMIPS RISC architecture]{
    \includegraphics[width=0.65\columnwidth]{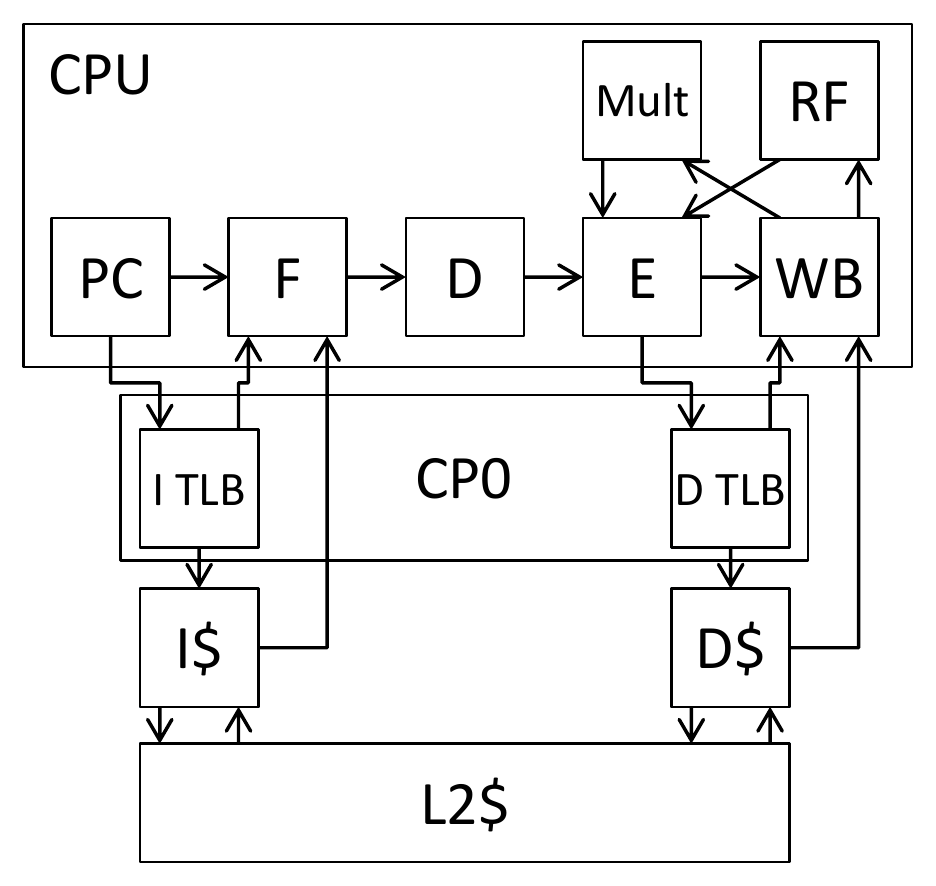}
  }
  \caption{The two example models analyzed.}
  \label{fg:examples}
\end{figure}

In this section we describe the architectures chosen to demonstrate our methodology and the results obtained using our tools. The two testcases are:

\textbf{Reed-Solomon:} This design is a parameterized Reed-Solomon error correction decoder which meets the throughput requirement for use in an 802.16 wireless receiver. The decoding algorithm is composed of several steps, each of which is implemented as a separate module shown in Figure~\ref{fg:rs_arch}. Dynamic activity, used for determining the power score metrics of the design, was generated using a testbench that feeds input data with errors at 50\% of the maximal correctable rate.


\textbf{SMIPS:} This design is a 32-bit RISC microarchitecture that implements the MIPS I ISA. Figure~\ref{fg:smips_arch} shows the main components consisting of a multiply unit, coprocessor 0 (implementing data and instruction TLBs), independent instruction/data L1 caches, and a unified, N-way L2 cache. This 5-stage processor can boot the GNU/Linux kernel. We used the dynamic activity generated during booting of Linux to generate the power score metrics of the SMIPS design.

We implemented these designs on three different FPGA devices: Spartan 6 XC6SLX45T-3FGG484, Virtex 5 XC5VLX155T-2FF1136 and Virtex 7 XC7VX485T-2FFG1761. We used the Xilinx ISE and Vivado tools 14.4 to synthesize, place and route the Verilog hardware model. We also used these tools to export the EDIF and SDF models. All the designs were targeted at 100 MHz on all the FPGAs. 



\subsection{Discussion of the results}

\begin{figure*}[t!]
  \centering
  \subfloat[Reed-Solomon area, delay and power metrics.\label{fg:rs_charts}]{
    \includegraphics[width=\textwidth]{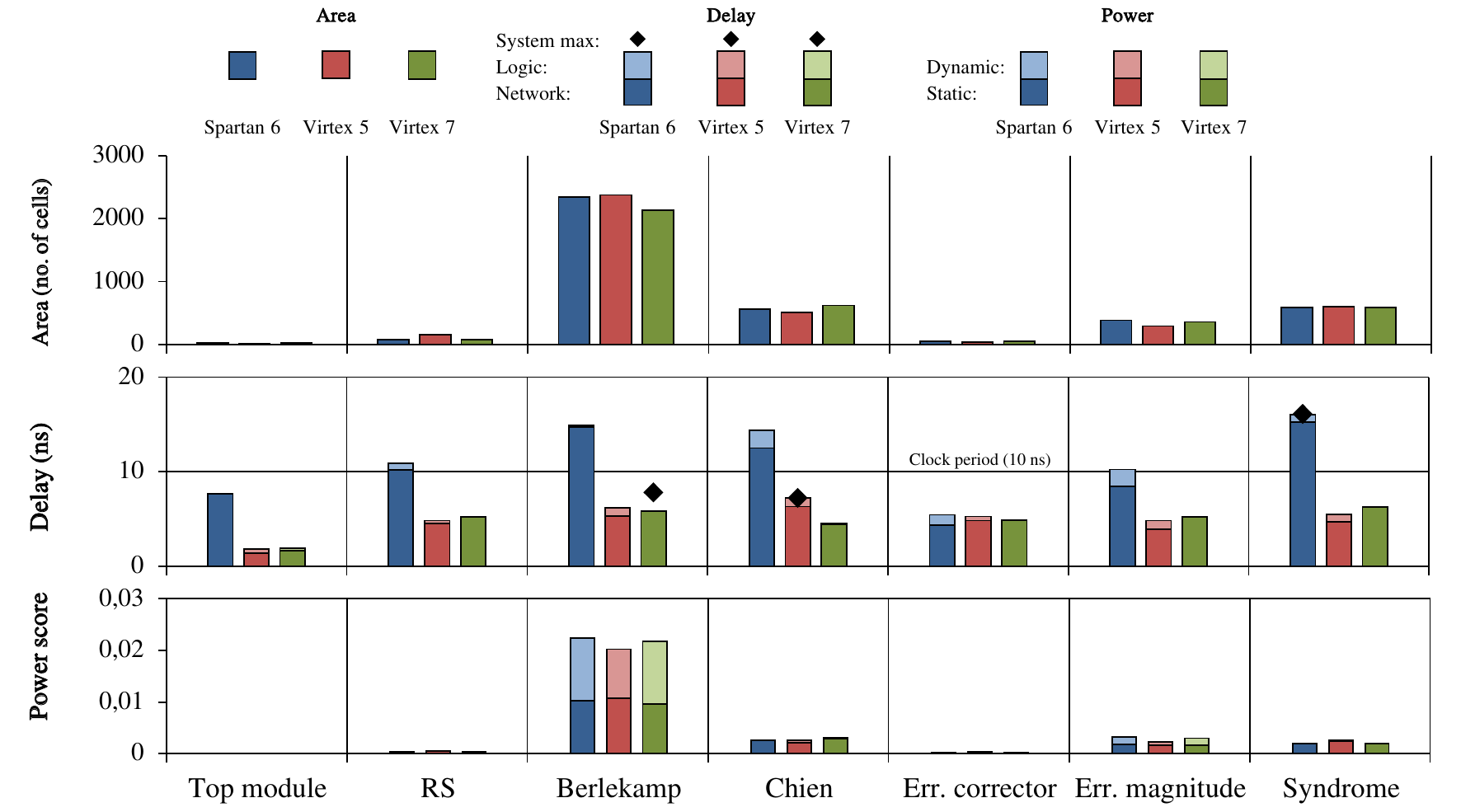}
  }
  \\
  \subfloat[SMIPS area, delay and power metrics.\label{fg:smips_charts}]{
    \includegraphics[width=\textwidth]{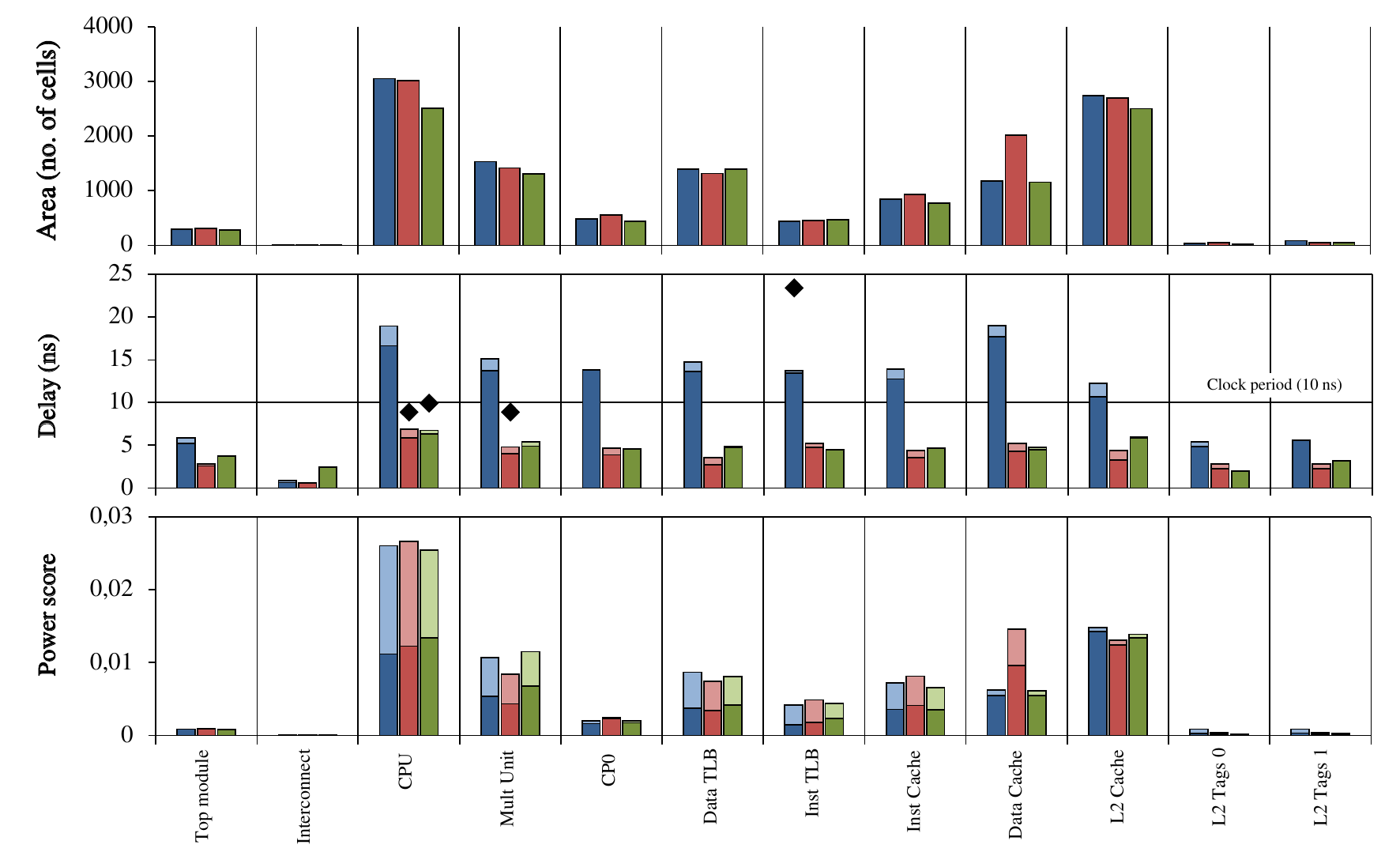}
  }
  \caption{Area, delay and power metrics on three Xilinx FPGAs: Spartan 6 (blue), Virtex 5 (red) and Virtex 7 (green).}\label{fg:charts}
\end{figure*}

The results for both architectures are shown in Figure~\ref{fg:charts}. The bar colors indicate the FPGA device: Spartan 6 -- blue, Virtex 5 -- red, and Virtex 7 -- green. Three metrics are displayed:

\begin{itemize}
\item The area results show the number of cells per architectural module.
\item The delay charts show the longest block delay, as previously explained in section \ref{sec:delay}. This metric measures the maximum contribution of each block to any path that crosses it. The darker components are the network part of the delay, and the lighter are the logic part. The diamond-shaped mark over the columns indicates the maximum delay of the system, and which block or blocks contribute to it. 
For both designs we can observe the significance of the network in the total delay. 
\item The power scores are also decomposed in two: static power score (darker shade) and dynamic power score (lighter shade).
\end{itemize}

Our framework produces results for every module and sub-module, but we grouped some results to simplify the charts.

\subsubsection{Reed-Solomon decoder}
Figure~\ref{fg:rs_charts} shows the breakdown of metrics for modules of the algorithm, as well as for the decoder module (RS) itself, which acts as a wrapper around these component modules, and for the top module which deals with Input-Output to memory. From an algorithmic perspective, Berlekamp step is the most computationally intensive part of the decoding process, and accordingly we see that this module has the maximum area in all three FPGA platforms. Error correction step involves the minimum computation as it simply removes the computed error values from the received data, thus contributing to minimal area. The other three component modules have similar moderate area usage. 

For delay metrics, implementations on Virtex 5 and Virtex 7 platforms easily meet the required 100 MHz clock frequency with critical paths located mostly in Chien and Berlekamp modules respectively. However, the Spartan 6 implementation is unable to achieve this due to long computational operations in Berlekamp, Chien and Syndrome, with Syndrome contributing the critical path. The power metrics roughly track in similar ratios as with the area metrics. One important point to notice is that most of the dynamic power consumption in the decoder is contributed by the Berlekamp module. Dynamic power comprises up to 50\% of Berlekamp's power consumption (in the case of Virtex 5) while other blocks' power consumption is mainly the static power of the FPGA resources used. This highlights the importance of this module for the decoder design, and suggests design refinement for reduced area as well as the use of power reduction techniques for reducing unnecessary dynamic activity (\eg clock gating).





\subsubsection{SMIPS}
In SMIPS there are 12 modules. These modules correspond to some of the architectural units shown in Figure~\ref{fg:smips_arch}. The results of the missing submodule architectural units, such as the pipeline stages, are included in their parent modules.
As with the previous case, the area and power metrics are very similar across all tested FPGA platforms, seen in Figure~\ref{fg:smips_charts}. 
In general, area metrics seem to follow a descending trend from Spartan 6 to Virtex 7. This is a result of the FPGA architectures being different in these devices. For instance, Spartan 6 slices (a group of two LUTs) have one carry chain output~\cite{Spartan6UG}, which can be used to implement fast carry chain arithmetic operations. The Virtex 5 and Virtex 7 slices have two independent carry chains~\cite{Virtex5UG}, which allow implementing more arithmetic operations with fewer LUTs. This is especially clear in the Virtex 7 area results.

The data cache requires more resources in Virtex 5 than in the other devices. The area report showed that the data cache module required 500 more registers in Virtex 5. We observed that while specifying the same architecture, the synthesis tool did not infer the data cache RAM unit correctly for Virtex 5, and implemented the cache memory using registers instead of using efficient on-chip BlockRAMs.
We argue that such portability problems make necessary not only the development of platform-neutral synthesis tools and languages, but also cross-platform analysis tools like ours.

The delay results differ for Spartan 6, which was unable to achieve timing closure for a target clock period of 10 ns. The area of the design has an important impact on the performance of the design. 
High resource usage congests the network and makes it difficult for the router to achieve the timing goals. SMIPS occupies about 30\% of the Spartan 6 device, much higher than the other two devices. In a congested device the network delays are high, even if it can fit the design. In addition, the logic delays of Spartan 6 cells are higher than the high-performance Virtex 5 and 7 LUTs. For instance, delay of a 6-input LUT in a Spartan 6 device can be $\sim$200~ps, whereas in Virtex 5 it is $\sim$80~ps and in Virtex 7 $\sim$40~ps. We can observe that the critical path of the Spartan 6 implementation is caused by the instruction TLB. In Virtex 5, we show two maximum delay marks, one over the CPU and another over the multiply unit. This means that the critical delay starts at the execution stage of the CPU and ends at the multiply unit. The delay report, along with the delay value, also includes the path that caused it and which architectural elements contribute to it. In Virtex 7, the maximum delay is caused by the execution stage of the CPU.

For power consumption,
it is seen that similar blocks dominate in all three platforms. These are L2 Cache, Execute block and the Multiplier. The dominance of the L2 Cache comes due to it being the largest block by far, thus having the largest static power dissipation. The computationally intensive Execute and Multiply blocks have a lot of logic and see a lot of dynamic activity. Beyond these three blocks we start seeing differences between the platforms. Virtex 5 has the Decode unit at relatively higher power consumption than even the Multiply unit. These differences arise due to the different availability of DSP arithmetic resources in the 3 FPGAs, different number of multiplexers generated for large data storage, and different levels of power and area optimizations implemented in the platforms. 


\section{Automatic Architectural Optimization}
\label{sec:archopt}

The architecture defined by the user determines the performance, area and power consumption of the final FPGA circuit. The way that the architecture is synthesized, placed and routed can optimize these metrics, but they are always constrained by the architectural decisions. Thus, we believe that significant changes of these results can only be achieved through high-level, architectural decisions. For instance, the results in Figure~\ref{fg:smips_charts} suggest several modifications at architectural level: reducing the number of entries of the L2 cache can improve area and power metrics. Splitting the Execute stage of the pipeline in two would break the combinational path crossing the data TLB and the data cache.

Currently these architectural optimizations are performed by the designer, under the guidance of the reports produced by the synthesis tools. Like the designer, our framework has knowledge about the architectural design and the synthesis reports. This knowledge enables the tool to implement technology-guided architectural changes.

The quality and impact evaluation methodology that we present in this work relies on two fundamental components. One is describing the hardware architecture using a high-level design language, such as Bluespec or another HLS language. The other is the methodology to project the technology problems to the architecture, as we described in the previous sections. But automatic architectural optimization requires additional components. The framework must distinguish the characteristics of each architectural unit, so that these parameters can be modified to meet the constraints imposed by the technology. For instance, the optimization tool should be able to modify the cache policies or the size of some units. The user should be able to put some constraints over those variations, informing what quality minimums must be preserved when modifying the architecture. Information about the target technology can complement these optimization inputs, allowing the tool to apply different strategies.


\section{Generalization of the Methodology and Related Work}
\label{sec:applicability}

We have shown how to implement our methodology for rule-based languages. In this section we want to discuss the generality of this method, and how it could be applied to HLS languages and tools.
Popular HLS tools such as xPilot~\cite{chen2005xpilot} (later AutoPilot and currently Xilinx Vivado~\cite{Vivado:www}), LegUp~\cite{canis2011legup} and ROCCC~\cite{villarreal2010designing} convert C programs into synthesizable hardware. These three examples use LLVM as their C front-end. The intermediate representation is then converted into Verilog or VHDL descriptions. 
The hardware architecture produced by C-based HLS converts blocks of C code, \ie functions, into Finite State Machines (FSM). Local and global variables are stored in local (block RAM) or external (DDR) memories. The first step of our methodology, the architectural analysis and annotations, requires augmenting the intermediate representation with architectural information. This can be applied to C programs reusing popular debugging information
that most compilers, including LLVM, support (for instance, DWARF for ELF files). 

In contrast to RTL or rule-based models, the FSMs generated by HLS require a variable number of cycles to finish not directly known at design time.
But the FSMs have start and finish signals. In this sense, performance analysis is similar to the implementation previously shown. In both cases the product is an RTL hardware description, where timing delays can be analyzed and tracked to the original functional blocks. The power analysis is essentially the same, C-based HLS also requiring functional simulation to obtain the execution rates of the architectural blocks. 

Regarding related work, Yan \etal \cite{Yan:AreaDelay} presented an estimation model that provides an area-delay tradeoff for chosen applications and FPGA platforms. However, it is aimed primarily at design partitioning of VLIW and Coarse-Grained reconfigurable architectures, while our work aims at modeling any custom hardware design. Modeling frameworks like McPAT~\cite{McPat09} are able to estimate design metrics for a wide variety of processor configurations and implementation technologies, but are limited to pre-defined architectural parameters and can not be used on arbitrary designs. Amouri~\etal~\cite{Amouri13} proposed a method to accurately measure and validate the leakage power distribution in FPGA chips using a thermal camera. These extremely accurate results can be used within our methodology for modeling architectural power consumption.
Li~\etal~\cite{Li05powermodeling} proposed a fine-grained power model for interconnects and LUTs in an FPGA implementation targeting sub-100 nm technology. However, correlating high-level design blocks to the FPGA power estimates requires additional analysis to keep track of how resources are allocated in each synthesis, placement and routing process, as well as individual activity and trace generation for various component blocks. Our technique provides this analysis.

\section{Conclusion and Future Work}
\label{sec:conc}

The increasing use of high-level hardware design languages is enlarging the gap between target technology and architectural specifications. Hardware designers require relevant feedback from post-synthesis tools that inform design decisions in an iterative process. \textbf{In this paper, we describe a methodology to relate post-synthesis area, delay and power data back to the initial HLS design.}
This methodology is a novel approach to architecture characterization. Unlike other techniques, it does not need additional user input to analyze the architecture. Instead, it uses the same hardware description used to synthesize the final circuit. The quality of the characteristics extracted from this circuit are backed by the quality of the FPGA vendor's tools. Such feedback allows the designer to quickly gauge the impact of architectural decisions on the quality of generated hardware.

At present, the design changes necessitated by the design constraints and the feedback generated using our technique have to be manually done by the user. Automation of design changes requires appropriate granularity in quantifying the impact of changes on area, performance and power metrics. \textbf{By satisfying this need, the work presented in this paper can serve as a foundation for Automatic Architectural Optimization.} We will investigate this possibility in the future. 

To summarize, we have implemented a tool that automates design characterization analysis and shown how it can help to improve the quality of final hardware and meet required goals. \textbf{For that purpose, we use two designs: a Reed-Solomon error correction decoder and a 32-bit pipelined processor implementation. We implement and characterize these designs on three FPGA platforms: Spartan 6, Virtex 5 and Virtex 7.} We discuss the limitations of the analysis and the impact of the final technology on the design, and we show examples of how the information reported by the tool can help to spot architectural problems. \textbf{Finally, this work has a high potential for use in automatic architectural optimization and cross-platform characterization, and could be applied on other HLS design languages and tools}.

\vspace{-5pt}



\bibliographystyle{IEEEtran}
\bibliography{dhw}

\begin{thebibliography}{10}
\providecommand{\url}[1]{#1}
\csname url@samestyle\endcsname
\providecommand{\newblock}{\relax}
\providecommand{\bibinfo}[2]{#2}
\providecommand{\BIBentrySTDinterwordspacing}{\spaceskip=0pt\relax}
\providecommand{\BIBentryALTinterwordstretchfactor}{4}
\providecommand{\BIBentryALTinterwordspacing}{\spaceskip=\fontdimen2\font plus
\BIBentryALTinterwordstretchfactor\fontdimen3\font minus
  \fontdimen4\font\relax}
\providecommand{\BIBforeignlanguage}[2]{{%
\expandafter\ifx\csname l@#1\endcsname\relax
\typeout{** WARNING: IEEEtran.bst: No hyphenation pattern has been}%
\typeout{** loaded for the language `#1'. Using the pattern for}%
\typeout{** the default language instead.}%
\else
\language=\csname l@#1\endcsname
\fi
#2}}
\providecommand{\BIBdecl}{\relax}
\BIBdecl

\bibitem{Chisel12}
J.~Bachrach, H.~Vo, B.~Richards, Y.~Lee, A.~Waterman, R.~Avizienis,
  J.~Wawrzynek, and K.~Asanovic, ``{Chisel: constructing hardware in a Scala
  embedded language},'' in \emph{DAC}, 2012, pp. 1216--1225.

\bibitem{Bluespec:www}
{Bluespec Inc.}, ``{Bluespec SystemVerilog},'' \url{www.bluespec.com}.

\bibitem{canis2011legup}
A.~Canis, J.~Choi, M.~Aldham, V.~Zhang, A.~Kammoona, J.~H. Anderson, S.~Brown,
  and T.~Czajkowski, ``{LegUp: high-level synthesis for FPGA-based
  processor/accelerator systems},'' in \emph{{Proceedings of the 19th
  International Symposium on Field Programmable Gate Arrays (FPGA)}}.\hskip 1em
  plus 0.5em minus 0.4em\relax ACM, 2011.

\bibitem{chen2005xpilot}
D.~Chen, J.~Cong, Y.~Fan, G.~Han, W.~Jiang, and Z.~Zhang, ``{xPilot: A
  Platform-Based Behavioral Synthesis System},'' \emph{SRC TechCon}, 2005.

\bibitem{villarreal2010designing}
{Villarreal, Jason and Park, Adrian and Najjar, Walid and Halstead, Robert},
  ``{Designing modular hardware accelerators in C with ROCCC 2.0},'' in
  \emph{{18th IEEE Annual International Symposium on Field-Programmable Custom
  Computing Machines (FCCM)}}.\hskip 1em plus 0.5em minus 0.4em\relax IEEE,
  2010.

\bibitem{Benson12}
J.~Benson, R.~Cofell, C.~Frericks, C.-H. Ho, V.~Govindaraju, T.~Nowatzki, and
  K.~Sankaralingam, ``{Design, integration and implementation of the DySER
  hardware accelerator into OpenSPARC},'' in \emph{{IEEE 18th International
  Symposium on High Performance Computer Architecture (HPCA)}}, 2012.

\bibitem{Moussalli13}
R.~Moussalli, W.~Najjar, X.~Luo, and A.~Khan, ``{A High Throughput No-Stall
  Golomb-Rice Hardware Decoder},'' in \emph{{IEEE 21st Annual International
  Symposium on Field-Programmable Custom Computing Machines (FCCM)}}, 2013.

\bibitem{cornu2011hls}
A.~Cornu, S.~Derrien, and D.~Lavenier, ``Hls tools for fpga: Faster development
  with better performance,'' in \emph{Reconfigurable Computing: Architectures,
  Tools and Applications}.\hskip 1em plus 0.5em minus 0.4em\relax Springer,
  2011, pp. 67--78.

\bibitem{Li05powermodeling}
F.~Li, Y.~Lin, L.~He, D.~Chen, and J.~Cong, ``{Power Modeling and
  Characteristics of Field Programmable Gate Arrays},'' \emph{IEEE Trans.
  Computer-Aided Design of Integrated Circuits and Systems}, vol.~24, no.~11,
  pp. 1712--1724, 2005.

\bibitem{Spartan6UG}
{Xilinx Inc.}, ``{Spartan-6 FPGA Configurable Logic Block User Guide},''
  \url{http://www.xilinx.com/support/documentation/user_guides/ug384.pdf}.

\bibitem{Virtex5UG}
------, ``{Virtex-5 FPGA User Guide},''
  \url{http://www.xilinx.com/support/documentation/user_guides/ug190.pdf}.

\bibitem{Vivado:www}
------, ``{Vivado},''
  \url{http://www.xilinx.com/products/design-tools/vivado/}.

\bibitem{Yan:AreaDelay}
L.~Yan, T.~Srikanthan, and N.~Gang, ``Area and delay estimation for fpga
  implementation of coarse-grained reconfigurable architectures,'' in
  \emph{Proceedings of the 2006 ACM SIGPLAN/SIGBED conference on Language,
  compilers, and tool support for embedded systems}, ser. LCTES '06.\hskip 1em
  plus 0.5em minus 0.4em\relax New York, NY, USA: ACM, 2006, pp. 182--188.

\bibitem{McPat09}
S.~Li, J.-H. Ahn, R.~Strong, J.~Brockman, D.~Tullsen, and N.~Jouppi, ``{McPAT:
  An integrated power, area, and timing modeling framework for multicore and
  manycore architectures},'' in \emph{42nd Annual IEEE/ACM International
  Symposium on Microarchitecture (MICRO-42)}, 2009, pp. 469--480.

\bibitem{Amouri13}
A.~Amouri, H.~Amrouch, T.~Ebi, J.~Henkel, and M.~Tahoori, ``{Accurate
  Thermal-Profile Estimation and Validation for FPGA-Mapped Circuits},'' in
  \emph{{IEEE 21st Annual International Symposium on Field-Programmable Custom
  Computing Machines (FCCM)}}, 2013, pp. 57--60.

\end{thebibliography}
%



\end{document}